\documentclass[12pt]{article}

\usepackage{graphics,amssymb,epsfig,float}
\usepackage[usenames,dvips]{color}
\usepackage{graphicx}
\usepackage{epsfig}
\usepackage{rotating}
\usepackage{dcolumn}
\usepackage{bm}
\usepackage{cite}
\usepackage{amsmath}
\usepackage{setspace}

\def\lsim{\;\raise0.3ex\hbox{$<$\kern-0.75em\raise-1.1ex\hbox{$\sim$}}\;}
\def\gsim{\;\raise0.3ex\hbox{$>$\kern-0.75em\raise-1.1ex\hbox{$\sim$}}\;}
\def\beq{\begin{equation}}   \def\eeq{\end{equation}}
\def\ba{\begin{array}}       \def\ea{\end{array}}
\def\bea{\begin{eqnarray}}   \def\eea{\end{eqnarray}}

\def\nl{\newline}
\def\k{\kappa}
\def\l{\lambda}

\def\noi{\noindent}

\begin{document}

\begin{titlepage}
\begin{flushright}
LPT Orsay 11-56\\
PCCF RI 1104
\end{flushright}

\begin{center}
\vspace{1cm}
{\Large\bf NMSDECAY: A Fortran Code for Supersymmetric Particle Decays
in the Next-to-Minimal Supersymmetric Standard Model}
\vspace{2cm}

{\bf{Debottam Das$^1$, Ulrich Ellwanger$^1$ and Ana M. Teixeira$^2$}}
\vspace{1cm}\\
\it $^1$  Laboratoire de Physique Th\'eorique, UMR 8627, CNRS and
Universit\'e de Paris--Sud,\\
\it B\^at. 210, 91405 Orsay, France \\
\it $^2$ Laboratoire de Physique Corpusculaire, CNRS/IN2P3 -- UMR
6533,\\
\it Campus des C\'ezeaux, 24 Av. des Landais, F-63171 Aubi\`ere Cedex,
France\\

\end{center}

\vspace{1cm}

\begin{abstract}
The code NMSDECAY allows to compute widths and branching ratios of
sparticle decays in the Next-to-Minimal Supersymmetric Standard Model.
It is based on a generalization of SDECAY, to include the 
extended Higgs and neutralino sectors of the NMSSM. 
Slepton 3-body decays, 
possibly relevant in the case of a singlino-like 
lightest supersymmetric particle,
have been added. NMSDECAY will be part of the NMSSMTools package,
which computes Higgs, sparticle masses and Higgs decays
in the NMSSM.
\end{abstract}

\end{titlepage}

\newpage
\section{Introduction}

The search for supersymmetric particles (sparticles) is one of the most
exciting tasks of present and future experiments in particle physics,
notably at the LHC \cite{Ball:2007zza,Aad:2009wy}. These searches can
only be successful if the possible signals of sparticles are known so
that they can be simulated and be compared to the data.  Sparticle
signals are not unique, since they depend on the specific supersymmetric
(SUSY) extension of the Standard Model (SM), i.e., on the particle
content, couplings and spectrum, which determine the numerous possible
decay cascades.

In recent years, SUSY decay cascades have been studied
mostly in the framework of the Minimal Supersymmetric Standard Model
(MSSM). In order to study the dependence of the decay cascades on the
sparticle masses and couplings, it is very helpful to use numerical
programs which perform the calculation of the large number of 
sparticle widths and possible branching ratios. The sparticle masses and
couplings depend, in turn, on the parameters of the underlying
Lagrangian. Therefore it is reasonable to interface programs for the
sparticle decays with programs which compute the sparticle masses and
couplings (so-called spectrum calculators).

In the case of the MSSM, one such interface is the program SUSY-HIT
(SUspect-SdecaY-Hdecay-InTerface)~\cite{Djouadi:2006bz}. It uses the
program Suspect~\cite{Djouadi:2002ze} for the calculation of the Higgs
and sparticle spectra, SDECAY~\cite{Muhlleitner:2003vg} for the
sparticle branching ratios, and HDECAY \cite{Djouadi:1997yw} for the
branching ratios of the Higgs sector. Another program
including simultaneously the calculation of sparticle spectra and decays
is SPheno~\cite{Porod:2003um,Porod:2011nf}.

However, the MSSM is not the only possible supersymmetric 
extension of the SM; it is just the one with a minimal Higgs sector,
which in this case contains two SU(2) doublets
$H_u$ (coupling to up-type quarks) and $H_d$ (coupling to down-type
quarks and charged leptons). The price to pay for such a minimal Higgs content
is a supersymmetric Higgs mass term $\mu$ in the MSSM superpotential, 
whose phenomenologically required order of magnitude (the
weak or the SUSY breaking scale) is difficult to
explain~\cite{Kim:1983dt}. The $\mu$-problem of the MSSM can be solved in the
Next-to-Minimal Supersymmetric Standard Model (NMSSM - for recent reviews
see~\cite{Maniatis:2009re,Ellwanger:2009dp}). The Higgs sector of the
NMSSM consists of two SU(2) doublets $H_u$ and $H_d$ (as in the MSSM),
to which a gauge singlet superfield $S$ is added. Due to a $\l S H_u
H_d$ coupling in the superpotential, the vacuum expectation value (vev)
of $S$, $s$, generates a supersymmetric mass term $\mu_\mathrm{eff}=\l s$ of
the desired order of magnitude for $H_u$ and $H_d$. In its simplest
$Z_3$ invariant version, the superpotential of the NMSSM is scale
invariant; it is in fact the simplest phenomenologically acceptable
supersymmetric extension of the SM with this property.

The new superfield $S$ implies various additional physical states
when compared to the MSSM: a neutral CP-even Higgs boson, a CP-odd Higgs
boson and a neutralino. These states can be relatively light; if the
additional NMSSM-specific CP-even Higgs boson mixes very little with the
Higgs bosons $H_u$ and $H_d$ of the MSSM, it will have very small
couplings to SM
gauge bosons, quarks and leptons and is hence much less 
constrained by the so far unsuccessful direct searches, 
notably at LEP~\cite{Schael:2006cr}. 
In some regions of the parameter space of the NMSSM,
CP-odd Higgs bosons can also be lighter than in the 
MSSM~\cite{Dobrescu:2000jt,Dobrescu:2000yn}. 
If they are light enough, any of the NMSSM-specific
additional Higgs and neutralino states can have an important impact on
sparticle decays, both as potential final states and/or as virtual
intermediate states in 3-body decays. Notably in the case of the 
constrained NMSSM (with universal soft SUSY breaking terms at the 
GUT scale~\cite{Djouadi:2008yj,Djouadi:2008uj}), a very weakly coupled
singlino-like neutralino can be the lightest supersymmetric particle
(LSP), thus affecting the final states of all sparticle decay cascades.

The impact of these NMSSM specific features on sparticle production
and on cascade decays has already been addressed:
sparticle decays into singlino-like neutralinos have
been studied in~\cite{Ellwanger:1997jj,Ellwanger:1998vi, Martin:2000eq,
Hesselbach:2000qw,Barger:2006kt,Djouadi:2008uj}, sparticle decays into
neutralinos in~\cite{Choi:2004zx}, decays of stops, sbottoms and
staus in~\cite{Kraml:2005nx}, the associated production of a light
CP-odd Higgs boson with a chargino pair in~\cite{Arhrib:2006sx},
neutralino decays into light Higgs bosons in~\cite{Cheung:2008rh},
neutralino decays into soft leptons in~\cite{Kraml:2008zr}, and
neutralino and chargino decays into many leptons in~\cite{Barger:2010aq}.

The purpose of the code NMSDECAY is to compute all sparticle
decays into all possible 2- and 3-body final states in the NMSSM:
squarks, sleptons, gluino, neutralinos and charginos decaying into
squarks, sleptons, gluino, neutralinos, charginos plus Higgs bosons,
SM-like quarks, leptons and gauge bosons. R-parity is assumed to be
conserved such that each sparticle can only decay into a lighter
sparticle plus SM-like particles and/or Higgs bosons.

NMSDECAY is based on an extension of the MSSM-specific code 
SDECAY~\cite{Muhlleitner:2003vg}, generalized to include 
the NMSSM-specific extensions of the
Higgs and neutralino sectors. QCD corrections, loop-induced and 3-body
decays are included as in SDECAY. 
In the NMSSM, a singlino-like LSP can lead to
dominant 3-body decays of sleptons, even if their 2-body decays are
kinematically possible. Such 3-body decays of sleptons (which were omitted in
SDECAY) have been added to NMSDECAY.

NMSDECAY is interfaced with NMSSMTools\footnote{Available on the web
page {\sf http://www.th.u-psud.fr/NMHDECAY/nmssmtools.html}}, and is
also written in FORTRAN~77. The installation and compilation of
NMSSMTools is a fairly straightforward procedure, 
described in the manual README, which can be found 
on the web site. The package NMSSMTools
includes: NMHDECAY~\cite{Ellwanger:2004xm,Ellwanger:2005dv}, where the
spectrum for the general NMSSM is computed and Higgs decays are
calculated similar to HDECAY~\cite{Djouadi:1997yw}; 
NMSPEC~\cite{Ellwanger:2006rn} where universal soft SUSY breaking terms at the
GUT scale are assumed; 
a link to MicrOMEGAS~\cite{Belanger:2006is,Belanger:2008sj}, which 
allows for the calculation of the dark matter relic
density and its detection cross sections. Sparticle decays of the NMSSM
with gauge mediated SUSY breaking \cite{Ellwanger:2008py} are not yet
included in NMSDECAY. Once NMSSMTools is compiled, the subroutines
required for NMSDECAY are linked automatically to NMHDECAY and NMSPEC.
However, the sparticle total widths and branching ratios are computed
and written only if a corresponding flag in the input file 
({\sf switch 13}) is switched on.

In the next section we give the Lagrangian and the particle content of
the NMSSM, and in Section~3 we describe how NMSDECAY is linked to
NMSSMTools, as well as the various calculations performed in the
different
subroutines. In the Appendix we describe the arguments of the
subroutines and {\sf COMMON} statements, which have to be defined if NMSDECAY
is called from another spectrum calculator.

\section{Lagrangian and Particle Content of the NMSSM}

The NMSSM differs from the MSSM due to the presence of the gauge singlet
superfield $S$. In the simplest $Z_3$ invariant realisation of the
NMSSM, the Higgs mass term $\mu H_u H_d$ in the superpotential
$W_{MSSM}$ of the MSSM is replaced by the coupling $\lambda$ of $S$
to $H_u$ and $H_d$, and a self-coupling $\kappa S^3$. 
Hence, in this simplest version the
superpotential $W_{NMSSM}$ is scale invariant, and given by:
\beq\label{eq:1}
W_{NMSSM} = \lambda \hat S \hat H_u\cdot \hat H_d + \frac{\kappa}{3} 
\hat S^3 +h_t \hat Q\cdot \hat H_u\, \hat T^c_R
+ h_b \hat H_d\cdot \hat Q\, \hat B^c_R + 
h_\tau \hat H_d \cdot \hat L\, \hat \tau^c_R\; ,
\eeq
where we have confined ourselves to the Yukawa couplings of $H_u$ and
$H_d$ to the quarks $Q$, $T_R$, $B_R$ and leptons $L$, $\tau_R$ of the
third generation. A sum over all generations is implicitly assumed. 
Once the scalar component of $S$ develops a vev $s$, 
the first term in $W_{NMSSM}$ generates an effective $\mu$-term
\beq\label{eq:2}
\mu_\mathrm{eff}=\lambda s\; .
\eeq

The soft SUSY breaking terms consist of mass terms for the
gauginos $\tilde{B}$ (bino), $\tilde{W}^a$ (winos) and $\tilde{G}^a$
(gluinos)
 \beq\label{eq:3}
-{\cal L}_\mathrm{\frac12}= \frac{1}{2} \bigg[ 
 M_1 \tilde{B}  \tilde{B}
\!+\!M_2 \sum_{a=1}^3 \tilde{W}^a \tilde{W}_a 
\!+\!M_3 \sum_{a=1}^8 \tilde{G}^a \tilde{G}_a   
\bigg]+ \mathrm{h.c.}\;,
\eeq
as well as soft breaking terms
for the Higgs bosons $H_u$, $H_d$ and $S$, squarks
$\tilde{q} \equiv (\tilde{t}_L, \tilde{b}_L$), $\tilde{t}_R$,
$\tilde{b}_R$ and sleptons $\tilde{\ell} \equiv (\tilde{\nu}_L,
\tilde{\tau}_L$) and $\tilde{\tau}_R$ 
(again in a simplified notation corresponding only to the third generation),
\beq\label{eq:4}
 -{\cal L}_\mathrm{0} =
m_{H_u}^2 | H_u |^2 + m_{H_d}^2 | H_d |^2 + 
m_{S}^2 | S |^2 +m_{\tilde{q}}^2|\tilde{q}^2| 
+ m_{\tilde{t}}^2|\tilde{t}_R^2|
+\,m_{\tilde{b}}^2|\tilde{b}_R^2| +m_{\tilde{\ell}}^2|\tilde{\ell}^2|
+m_{\tilde{\tau}}^2|\tilde{\tau}_R^2|\; .
\eeq

Finally there are trilinear interactions involving the sfermion and the
Higgs fields, also including the singlet field 
\beq
-{\cal L}_\mathrm{tril}= 
 \Bigl( h_t A_t\, Q\cdot H_u\, T_R^c + h_b  A_b\, H_d \cdot Q\, B_R^c +
h_\tau A_\tau \,H_d\cdot L \,\tau_R^c  +\,  \lambda A_\lambda\, H_u
\cdot H_d \,S +  \frac{1}{3} \kappa  A_\kappa\,  S^3 \Bigl)+
\mathrm{h.c.}\;.
 \label{eq:5}
\eeq

After electroweak symmetry breaking, and once the mass matrices 
have been diagonalized, the above fields give rise to
the following physical eigenstates:

\vspace*{3mm}
\noi - Higgs sector (assuming CP-conservation):\nl
\phantom{xxxx} 3 neutral CP-even Higgs bosons $H_i$, which are mixtures
of the CP-even components\nl
\phantom{xxxxxx} of the superfields $H_u$, $H_d$ and $S$;\nl
\phantom{xxxx} 2 neutral CP-odd Higgs bosons $A_i$, which are mixtures of
the CP-odd components\nl
\phantom{xxxxxx} of $H_u$, $H_d$ and $S$;\nl
\phantom{xxxx} 1 charged Higgs boson $H^\pm$, a mixture of the
charged components of $H_u$ and $H_d$.

\vspace*{3mm}
\noi - 5 neutralinos $\tilde{\chi}^0_i$, which are mixtures of the bino
$\tilde{B}$, the neutral wino $\tilde{W}^3$, the neutral higgsinos from
the superfields $H_u$ and $H_d$, and the singlino from the superfield
$S$.

\vspace*{3mm}
\noi - 2 charginos $\tilde{\chi}^\pm_i$, which are mixtures of the
charged winos $\tilde{W}^{\pm}$ and the charged higgsinos from the
superfields $H_u$ and $H_d$.

\vspace*{3mm}
\noi - Gluinos $\tilde{G}$, which correspond to physical eigenstates.

\vspace*{3mm}
\noi - 4 complex scalar quarks $\tilde{q}$ per generation, which can be
separated into up-type and down-type squarks.  For the first two
squark generations, mass eigenstates correspond to the weak
interaction eigenstates, which carry the
quantum numbers of the corresponding left- and right-handed quarks. The
squarks of the third generation have to be treated separately, since
mixings proportional to the Yukawa couplings are relevant: for the third
generation we have 2 stops $\tilde{t}_{1,2}$ (with $m_{\tilde{t}_{2}} >
m_{\tilde{t}_{1}}$) and 2 sbottoms $\tilde{b}_{1,2}$ (likewise ordered
in mass).

\vspace*{3mm}
\noi - 3 complex scalar leptons $\tilde{\ell}$ per generation, 2 of which
are charged and one is neutral. As in the case of squarks, the
slepton mass eigenstates of the first two generations correspond to the 
weak eigenstates with the
quantum numbers of the corresponding left- and right-handed leptons
(we only consider left-handed (s)neutrinos). For
the third generation we have 2 staus $\tilde{\tau}_{1,2}$ (ordered in
mass) and a sneutrino $\tilde{\nu}_\tau$.

\section{The Structure of NMSDECAY}
\subsection{Compilation, Input and Output}

From version NMSSMTools\_3.0.0 onwards, the required files for
NMSDECAY are included in the directory {\sf sources} of NMSSMTools. Once
the commands {\sf make init} and {\sf make} are typed in the main
directory, these files are compiled and linked automatically to the
routines {\sf nmhdecay} and {\sf nmspec}. The compilation is managed
by the routine {\sf getFlags} inside {\sf micromegas/CalcHEP\_src},
which checks automatically for the available compilers: the default for
Linux is {\sf gfortran}, but {\sf g77} and {\sf ifort} could also be
used. (In the latter case the corresponding flags in {\sf getFlags} have
to be set by the user.)

The input files have to be named according to the convention {\sf
  \#inp\#}, where the prefix and suffix of the input file
can be chosen at will by the user (at least one
of the {\sf \#} must be non-vanishing). The format of the input files
has to follow the conventions specified in the SUSY Les Houches
Accord~2~\cite{Allanach:2008qq}, and include a {\sf BLOCK MODSEL}
where the {\sf switch~3} (choice of particle content) must be set to~{\sf 3}
(NMSSM). In order to allow NMSDECAY to run,
{\sf switch~1} (choice of SUSY breaking model) must be set to {\sf 0}
(general NMSSM) or {\sf 1} (mSUGRA = NMSPEC). 
Additional switches inside the {\sf BLOCK MODSEL} specify the
activation of various subroutines, and the {\sf switch~13} must be set 
to~{\sf 1} in order to activate NMSDECAY. No additional action by the user is
required.

Once NMSDECAY is activated and the input file is run (typing {\sf (./) run
\#inp\#}), the sparticle decay widths and branching ratios are appended
to the output file {\sf \#decay\#}, where the prefix and suffix correspond
to the name of the input file. (This file always contains the various
Higgs decay widths and branching ratios.) The format of the output also
follows the SUSY Les Houches Accord \cite{Skands:2003cj}, with
particle codes for the NMSSM as specified in \cite{Allanach:2008qq}.
If scans are activated, the user has to define the format of the output
in the subroutines {\sf OUTPUT} using the branching ratios made available in the
various {\sf COMMON} statements (see Section~3.3).

\subsection{Contents of {\sf NMSDECAY\_INTERFACE}}

The {\sf main} routines in the files {\sf main/nmhdecay.f} and {\sf
main/nmspec.f} only call the subroutine {\sf NMSDECAY\_INTERFACE} in the
file {\sf sources/NMSDECAY.f}. The subroutine {\sf
NMS\-DECAY\_\-INTERFACE} serves to:
(i) translate parameters, masses, mixing
angles and couplings from NMSSMTools into the conventions used in the
other sparticle-specific subroutines; (ii) call these subroutines; 
(iii) call the subroutine {\sf NS\_output}.

At the beginning of {\sf NMSDECAY\_INTERFACE}, various flags are set
to~{\sf 1}: {\sf flagqcd} for QCD corrections to 2-body decays, and {\sf
flagmulti} and {\sf flagloop} for the activation of 3-body and
loop-induced decays, respectively. In case the user is not interested in QCD
corrections and/or the latter decays (e.g. in order to speed up the running),
these flags can be set to~{\sf 0}.
Then, {\sf make init} and {\sf make} must be typed again for
re-compilation.

The parameter {\sf multilim} (default: {\sf multilim=0.01}) specifies
under which condition 3-body branching ratios are taken into account
({\it after} they have been computed). Only if a the total 3-body decay
width is larger than {\sf multilim} times the total 2-body decay width,
3-body branching ratios will be included in the sparticle widths and
in the output file. (Otherwise a dominant part of the output file can
contain a lengthy list of very small 3-body branching ratios, which adds
little information.)

The QCD corrections depend on the scale $Q^2$ at which the couplings and
masses (notably the strong coupling $\alpha_s$) of the tree
level vertices are defined. This scale ({\sf amuref}) is set equal to
the SUSY scale, computed in terms of the squark masses of the first 
two generations in the routine {\sf sources/runpar}. Alternatively, the
SUSY scale can be specified in the input file. Note that various input
parameters are implicitely defined at the SUSY scale. Hence, a change of
$Q^2$ in the input file will lead implicitely to a different choice for
these input parameters (SUSY breaking terms and NMSSM-specific
parameters which affect, in turn, masses and mixing angles), and not
just to a change of scale of the couplings of the tree level vertices.
In any case, the $Q^2$ dependence of the QCD corrections of the 2-body
decays involving strongly interacting (s)particles compensates the
dominant $Q^2$ dependence of the tree level vertices, if the latter is
due to the running of the strong coupling constant.

Couplings, masses and mixing angles at the SUSY scale are read from
various {\sf COMMON} statements, which are initialised in various other
subroutines in NMSSMTools. For convenience, these {\sf COMMON}
statements - together with the arguments {\sf PAR} of the subroutine
{\sf NMSDECAY\_\-INTERFACE(PAR)} corresponding to NMSSM-specific input
parameters - are listed in the Appendix. The numerous couplings among
sparticles and SM-like particles are written into 
several NMSDECAY-specific {\sf COMMON} statements, which carry a 
{\sf NS\_} prefix.

\subsection{Contents of Sparticle Decay Files}

The sparticle decay files contain appropriate modifications of the
corresponding subroutines of SDECAY, generalizing the neutral Higgs and
neutralino sectors to include the particle content of the NMSSM.
For each species of sparticles, separate subroutines in the files {\sf
NS\_name\_of\_sparticle} are devoted to the calculation of the 2-body
and 3-body widths and branching ratios. For 2-body decays, 1-loop QCD
corrections have been included as in SDECAY (with the renormalisation
scale $Q^2$ given by the SUSY scale, unless modified by the user - see
above). In rare cases, 1-loop QCD corrections can be negative and larger
in absolute value than the tree level branching ratio, for any choice of
the renormalisation scale. Of course, this signals that higher order
corrections would be relevant. In order to avoid negative branching
ratios in such cases, we replace (provisionally) $(1+corr.)$ by
$1/(1-corr.)$ whenever the relative 1-loop correction $corr$ is less
than -1.

If 2-body decay channels are not kinematically open or are strongly
suppressed,
3-body decays can be relevant for sbottoms, stops, gluinos, neutralinos
and charginos. (Strong suppressions of 2-body decays can occur in the
case of a singlino-like LSP in the NMSSM.)
Then, the virtual exchanged particles can be neutral and charged Higgs
bosons, $W^\pm$, $Z$, gluino, charginos, neutralinos, squarks, sleptons
and the top quark. Interference terms for identical final states are
taken into account.

Since the quarks of the first two generations and the leptons are light,
in the MSSM
it is usually assumed that the corresponding $\tilde{q} \to q + 
\mathrm{LSP}$ and $\tilde{\ell} \to \ell + \mathrm{LSP}$ 2-body
decays are kinematically allowed and dominant, so that 3-body decays can be
safely neglegted. This is not always true in the NMSSM, even if one requires
a neutral LSP (for cosmological reasons): the LSP can be a singlino-like
neutralino ($\tilde \chi_1^0 \approx \tilde \chi_s^0$)
with very small couplings to squarks and sleptons. Notably
sleptons can be lighter than the lightest MSSM-like neutralino
($\tilde \chi_2^0$) such that
their decay into the singlino-like neutralino (plus the corresponding
lepton) would represent the only kinematically allowed 2-body decay
mode. Since this partial width would be very small, 3-body decays into
the lightest slepton (assumed to be the $\tilde{\tau}_1$) can be
dominant. These have been added to the decay modes already present in SDECAY.
(We take care of a subtlety in the case where 2-body decays are in
principle allowed, but subdominant: we ensure that a particle only
mediates a 3-body decay if its mass is sufficiently large to prevent it from
being produced in a 2-body decay; 
otherwise one would be generating double-countings in decay cascades, and
the phase space integrals would typically diverge.)

\vspace*{5mm}
In what follows we summarise the different {\sf NS\_} decay
subroutines, briefly describing the decays contained in each one: 
\vspace{3mm}

\noi {\sf NS\_squark.f}:\nl
\phantom{xxx} $\tilde{q}_{_{L,R}}$ 2-body decays (first two generations)
into $q+ \tilde{\chi}^0_i$, $q+ \tilde{\chi}^\pm_i$, $q+ \tilde{G}$.

\vspace{3mm}
\noi  {\sf NS\_sbottom.f}:\nl
\phantom{xxx} $\tilde{b}_{1,2}$ 2-body decays into $b+\tilde{\chi}^0_i$,
$b+ \tilde{G}$, $t+\tilde{\chi}^\pm_i$,
$\tilde{t}_{1,2}+H^\pm$, $\tilde{t}_{1,2}+W^\pm$;\nl
\phantom{xxx} $\tilde{b}_{2}$ 2-body decays into
$\tilde{b}_{1}+H_i$, $\tilde{b}_{1}+A_i$, $\tilde{b}_{1}+Z$;\nl
\phantom{xxx} $\tilde{b}_{1,2}$ 3-body decays into $t+\ell+\tilde{\ell^\prime}$,
$\tilde{t}_{1,2}+q+\bar{q^\prime}$, $\tilde{t}_{1,2}+\ell+\bar{\ell^\prime}$;\nl
\phantom{xxx} $\tilde{b}_{2}$ 3-body decays into
$\tilde{b}_{1}+q+\bar{q}$, $\tilde{b}_{1}+\ell+\bar{\ell}$.

\vspace*{3mm}
\noi  {\sf NS\_stop.f}:\nl
\phantom{xxx} $\tilde{t}_{1,2}$ 2-body decays into
$t+\tilde{\chi}^0_i$,
$t+ \tilde{G}$, $b+\tilde{\chi}^\pm_i$,
$\tilde{b}_{1,2}+H^\pm$, $\tilde{b}_{1,2}+W^\pm$;\nl
\phantom{xxx} $\tilde{t}_{2}$ 2-body decays into
$\tilde{t}_{1}+H_i$, $\tilde{t}_{1}+A_i$, $\tilde{t}_{1}+Z$;\nl
\phantom{xxx} Loop-induced $\tilde{t}_{1}$ decay into a charm quark
$+\ \tilde{\chi}^0_i$;\nl
\phantom{xxx} $\tilde{t}_{1,2}$ 3-body decays into $b+\ell+\tilde{\ell^\prime}$,
$\tilde{b}_{1,2}+q+\bar{q^\prime}$, $\tilde{b}_{1,2}+\ell+\bar{\ell^\prime}$;\nl
\phantom{xxx} $\tilde{t}_{2}$ 3-body decays into
$\tilde{t}_{1}+q+\bar{q}$, $\tilde{t}_{1}+\ell+\bar{\ell}$.

\newpage
\noi {\sf NS\_gluino.f}:\nl
\phantom{xxx} $\tilde{G}$ 2-body decays into $q + \tilde{q}_{_{L,R}}$ (all 3
generations);\nl
\phantom{xxx} Loop-induced $\tilde{G}$ decays into $g +
\tilde{\chi}^0_i$;\nl
\phantom{xxx} $\tilde{G}$ 3-body decays into $\tilde{\chi}^0_i + q +
\bar{q}$, $\tilde{\chi}^\pm_i + q + \bar{q^\prime}$, $\tilde{t}_{1,2} + b +
H^\pm$, $\tilde{t}_{1,2} + b + W^\pm$.

\vspace*{3mm}
\noi {\sf NS\_slepton.f}:\nl
\phantom{xxx} $\tilde{\ell}_{_{L,R}}$ ($\tilde \nu_{_L}$) 2-body decays
(first two generations) into
$\ell \,(\nu) + \tilde{\chi}^0_i$, $\nu \,(\ell) + \tilde{\chi}^\pm_i$;\nl
\phantom{xxx} $\tilde{\tau}_{1,2}$ 2-body decays into
$\tau + \tilde{\chi}^0_i$, $\nu_\tau + \tilde{\chi}^\pm_i$, 
$\tilde{\nu}_\tau + H^\pm$,  $\tilde{\nu}_\tau +W^\pm$;\nl
\phantom{xxx} $\tilde{\nu}_\tau$ 2-body decays into
$\nu_\tau+ \tilde{\chi}^0_i$, $\tau + \tilde{\chi}^\pm_i$,
$\tilde{\tau}_{1,2} + H^\pm$, $\tilde{\tau}_{1,2} + W^\pm$;\nl
\phantom{xxx} $\tilde{\tau}_{2}$ 2-body decays into $\tilde{\tau}_{1} +
H_i$, $\tilde{\tau}_{1} + A_i$, $\tilde{\tau}_{1} + Z$;\nl
\phantom{xxx} $\tilde{\ell}_{_{L,R}}$ ($\tilde
\nu_{_L}$) 3-body decays (first two generations) into
$\tilde{\tau}_{1}+\tau + \ell\,(\nu)$, $\tilde{\tau}_{1}+ \nu_\tau
+\nu \,(\ell)$;\nl
\phantom{xxx} $\tilde{\tau}_{2}$ 3-body decays into $\tilde{\tau}_{1} +
\tau^+ + \tau^-$, $\tilde{\tau}_{1} +
\nu_\tau + \bar{\nu}_\tau$;\nl
\phantom{xxx} $\tilde{\nu}_\tau$ 3-body decay into $\tilde{\tau}_{1} +
\tau + \nu_\tau$.

\vspace*{3mm}
\noi {\sf NS\_neutralino.f}:\nl
\mbox{\phantom{xxx} $\tilde{\chi}^0_i$ 2-body decays into
$\tilde{\chi}^0_j + H_i$, $\tilde{\chi}^0_j + A_i$, $\tilde{\chi}^0_j +
Z$, $\tilde{\chi}^\pm_i + W^\pm$, $\tilde{\chi}^\pm_i + H^\pm$,
$\ell +\tilde{\ell}$, $\nu +\tilde{\nu}$,}\nl
\phantom{xxx}  $q + \tilde{q}$;\nl 
\phantom{xxx} Loop-induced $\tilde{\chi}^0_i$ decays into
$\tilde{\chi}^0_j + \gamma$;\nl
\phantom{xxx} $\tilde{\chi}^0_i$ 3-body decays into
$\tilde{\chi}^0_j + q + \bar{q}$, $\tilde{\chi}^0_j + \ell +
\bar{\ell}$, $\tilde{\chi}^0_j + \nu + \bar{\nu}$,
 $\tilde{\chi}^\pm_i + q + \bar{q^\prime}$, $\tilde{\chi}^\pm_i + \ell + \bar{\ell^\prime}$,
$\tilde{G} + q + \bar{q}$.

\vspace*{3mm}
\noi {\sf NS\_chargino.f}:\nl
\phantom{xxx} $\tilde{\chi}^\pm_i$ 2-body decays into
$\tilde{\chi}^0_i + W^\pm$, $\tilde{\chi}^0_i + H^\pm$,
$\ell +\tilde{\ell^\prime}$, $q + \tilde{q^\prime}_{_{L,R}}$;\nl 
\phantom{xxx} $\tilde{\chi}^\pm_2$ 2-body decays into
$\tilde{\chi}^\pm_1 + H_i$, $\tilde{\chi}^\pm_1 + A_i$,
$\tilde{\chi}^\pm_1 +
Z$;\nl 
\phantom{xxx} $\tilde{\chi}^\pm_i$ 3-body decays into
$\tilde{\chi}^0_i + q + \bar{q^\prime}$, $\tilde{\chi}^0_i + \ell + \bar{\ell^\prime}$,
$\tilde{G} + q + \bar{q^\prime}$;\nl
\phantom{xxx} $\tilde{\chi}^\pm_2$ 3-body decays into
$\tilde{\chi}^\pm_1 + q + \bar{q}$, $\tilde{\chi}^\pm_1 + \ell + \bar{\ell}$.

\vspace*{3mm}
We recall that as a consequence of the extended Higgs and neutralino sectors of the
NMSSM, in the above decays one now has 
$\tilde \chi_i^0$ ($i=1,...,5$), $H_i$ ($i=1,...,3$) and $A_i$ ($i=1,2$).
 
For every sparticle decay computed in {\sf NS\_sparticle.f}, the
various widths, 2-body and 3-body branching ratios are collected in {\sf
COMMON} statements {\sf COMMON/SPARTICLE\_\-WIDTH}, {\sf
COMMON/SPARTICLE\_BR\_2BD} and {\sf COMMON/SPARTICLE\_BR\_3BD},
respec\-ti\-vely. These {\sf COMMON} statements are used in the file {\sf
NS\_output.f}, which writes the total widths and partial branching
ratios into the user-defined output file {\sf \#decay\#} using the SUSY
Les Houches Accord~2 format \cite{Allanach:2008qq} (see above).

Additional auxiliary subroutines are contained in the file {\sf
NS\_auxfunc.f}, notably routines from SDECAY concerning QCD corrections.

\vspace*{5mm}
Working in the so-called NMSSM decoupling regime ($\l,\ \k \to 0$), we
have verified that all partial
widths and branching ratios of processes not involving NMSSM specific
neutralino or Higgs states coincide with the results of SDECAY for the
corresponding MSSM parameters.

\enlargethispage{1\baselineskip}
NMSSM specific results of NMSDECAY (partial widths and branching ratios) have
also been compared to results previously obtained in the literature,
whenever details of spectrum and decays were available. In
particular, we confronted our results for the following processes:
$\tilde{\chi}^0_2 \to \tilde{\chi}^0_1 + A_1$ 
to~\cite{Cheung:2008rh};
$\tilde{\chi}^0_2 \to \tilde{\chi}^0_1 + \ell^+ + \ell^-$ 
to~\cite{Kraml:2008zr};
$\tilde{\chi}^0_i$, $\tilde{\chi}^\pm_1$,
$\tilde{\ell}_{_{L,R}}$ ($\tilde \nu_{_L}$) decays into many leptons
to~\cite{Barger:2010aq}; in all cases we found reasonable agreement.
Furthermore we compared several widths with the results of
SPheno~\cite{Porod:2003um,Porod:2011nf}, finding a fair consistency in
most of the cases.

To conclude, the routines in NMSDECAY will make it easy to compute all
sparticle decay widths and branching ratios in the NMSSM, once the
sparticle and Higgs spectra are computed by one of the spectrum
calculators in NMSSMTools; it suffices to switch on a corresponding flag
in the input file. Various applications will be published in a separate
paper. 

Since NMSDECAY is based on SDECAY, any publication using NMSDECAY should
also refer explicitly to SDECAY~\cite{Muhlleitner:2003vg}.

\section*{Acknowledgements}

We are very grateful notably to M. M\"uhlleitner, A. Djouadi and Y.
Mambrini for the permission to use large parts of the code SDECAY. We
thank W. Porod for discussions regarding SPheno. D.~D. acknowledges
support from the CNRS, and U.~E. from the French ANR LFV-CPV-LHC.

\newpage

\section*{Appendix}

Here we list the variables which are transferred to NMSDECAY from other
routines in NMSSMTools. The {\sf COMMON} statements appear both in {\sf
NMSDECAY\_INTERFACE} and, occasionally, in the other subroutines.

\vspace*{3mm}\noi
Parameters {\sf PAR(I),\ I=1...25,} in the argument of {\sf
NMSDECAY\_INTERFACE(PAR)}: the\break NMSSM-specific couplings $\l$,
$\k$, $\tan(\beta)$, $\mu_\mathrm{eff}$, and all soft SUSY breaking
parameters as specified at the beginning of {\sf nmhdecay.f} (i.e.
without the Higgs mass terms). All parameters are assumed to be defined
at the SUSY scale $Q^2$, with the exception of $\tan(\beta)$ 
(which is defined at $M_Z$).

\vspace*{3mm}\noi
{\sf COMMON/RENSCALE/}: the SUSY scale $Q^2$ where the couplings and
soft SUSY breaking terms are defined

\vspace*{3mm}\noi
{\sf COMMON/SUSYMH/}: soft SUSY breaking Higgs mass terms at the SUSY
scale $Q^2$

\vspace*{3mm}\noi
{\sf COMMON/GAUGE/}: gauge couplings and $\sin^2(\theta_W)$ at the scale
$M_Z$

\vspace*{3mm}\noi
{\sf COMMON/SUSYCOUP/}: gauge and Yukawa couplings at the SUSY scale
$Q^2$

\vspace*{3mm}\noi
{\sf COMMON/SMSPEC/}: SM quark, lepton and electroweak gauge boson
pole masses

\vspace*{3mm}\noi
{\sf COMMON/HIGGSPEC/}: NMSSM Higgs pole masses and mixing angles

\vspace*{3mm}\noi
{\sf COMMON/SUSYSPEC/}: gluino, chargino and neutralino pole masses and
mixing angles

\vspace*{3mm}\noi
{\sf COMMON/SFSPEC/}: squark and slepton pole masses and mixing angles

\vspace*{3mm}\noi
{\sf COMMON/QHIGGS/}: Higgs vevs and $\tan(\beta)$ at the scale QSTSB
(the scale of stop/sbottom masses, close to $Q^2$ in general)

\end{document}